\documentclass[prd,aps,floats,superscriptaddress,nofootinbib,twocolumn,floatfix]{revtex4-1}
\usepackage{exscale}                  % correct scaling of math-symbols
\usepackage[intlimits]{amsmath}       % improved mathematical typesetting
\usepackage{amsfonts}
\usepackage{amssymb,amscd}
\usepackage{wrapfig}
\usepackage{graphicx}
\usepackage{array}
\usepackage{bbm}
\usepackage{multirow}
\usepackage[T1]{fontenc}
\usepackage{times}
\usepackage{subfigure}
\usepackage[x11names]{xcolor}
\usepackage{epstopdf}

\newcommand{\nf}{N_{f}\,}
\newcommand{\vf}{v_{F}\,}
\newcommand{\vd}{v_{\Delta}\,}

\newcommand{\beq}{\begin{equation}}
\newcommand{\eeq}{\end{equation}}

\begin{document}

\title{Critical scaling of finite temperature $\mbox{QED}_3$ 
in anisotropic space-time}

\author{Jacqueline~A.~Bonnet}
\affiliation{Institut f\"ur Theoretische Physik, 
 Universit\"at Giessen, 35392 Giessen, Germany}
\author{Christian~S.~Fischer}
\affiliation{Institut f\"ur Theoretische Physik, 
 Universit\"at Giessen, 35392 Giessen, Germany}
\affiliation{GSI Helmholtzzentrum f\"ur Schwerionenforschung GmbH, 
  Planckstr. 1  D-64291 Darmstadt, Germany.}

\begin{abstract}
We investigate the scaling behavior of the critical temperature of
anisotropic QED in 2+1 dimensions with respect to a variation of the
number of fermions $N_f$. To this end we determine the order parameter 
of the chiral transition of the theory from a set of (truncated) 
Dyson--Schwinger equations for the fermion propagator formulated
in a finite volume. We verify the validity of previously determined 
universal power laws for the scaling behavior of the critical 
temperature with $N_f$. We furthermore study the variation of the 
corresponding critical exponent with the degree of anisotropy and find
considerable variations. 
\end{abstract}

\maketitle

%%%%%%%%%%%%%%%%%%%%%%%%%%%%%%%%%%%%%%%%%%%%%%%%%%%%%%%%%%%%%%%%%%%%%%%%%%%%%
\section{Introduction \label{sec:introduction}}

The properties of asymptotically free quantum field theories under variation 
of the number of fermion flavors are currently investigated with great efforts.
In particular the corresponding quantum and thermal phase transitions from a 
chirally symmetric to a conformal phase are attracting a lot of interest.
Within QCD, these efforts are triggered by the desire to understand the
origin and size of the conformal window and the associated potential
for applications for physics beyond the standard model, see 
Refs.~\cite{arXiv:0911.0931,arXiv:1111.2317,arXiv:1111.1575} and references 
therein. Another asymptotically free theory is QED in two space and one time
dimensions. QED$_3$ in its strong interaction version serves as a laboratory to 
explore theoretical concepts in the comparably simple Abelian framework. On the
other hand, QED$_3$ also has important applications as a potential effective 
theory for strongly interacting fermionic systems like graphene \cite{Novoselov:2005kj}
and high temperature cuprate superconductors \cite{Franz:2002qy,Herbut:2002yq}.
These superconductors are characterized by d-wave symmetry of the gap
function. The variation of the number of fermion flavors in these materials 
is tied to the number of two-dimensional conduction layers. Another particularly 
interesting feature of these materials is that they require a difference between  
the Fermi velocity, $v_F$, and a second velocity $\vd$ related to the amplitude 
of the superconducting order parameter. This fermionic anisotropy can be as large 
as \cite{Chiao:2000} $\lambda = \vf / \vd \sim 10$. Furthermore, both velocities
are also different from the speed of light, $c_s$. Thus, in the effective QED$_3$
theory all three Euclidean directions are different, leading to an anisotropic 
formulation of the theory. Consequently, the critical number of fermion flavors 
$N_f^c$ does depend on $(c_s,\vf,\vd)$, see 
Refs.~\cite{Lee:2002qza,Hands:2004ex,Thomas:2006bj,Concha:2009zj,Bonnet:2011hh} 
for recent studies. The theoretical study of this situation sheds light on the 
applicability of QED$_3$ to the situation in the superconductor but also offers the 
possibility to explore anisotropy effects in general with potential applications
to a description of the quark-gluon plasma in high temperature QCD. 

It is therefore also of great interest to introduce finite temperature into the
anisotropic QED$_3$ framework. Then, in addition to
the quantum phase transition at a zero temperature critical $N_f^c$,
one expects thermal transitions for fixed $N_f$ when the temperature 
reaches an $N_f$-dependent critical value. Particularly interesting is the
region where the thermal transition line merges into the quantum transition 
point. From very general arguments, universal scaling laws for such a situation
have been derived in Refs.~\cite{arXiv:0912.4168,Braun:2010qs}, see also
Ref.~\cite{Braun:2011pp} for a review. One finds
\beq
 T_{cr} \sim k_0 |N^c_{f,0} - N_f |^{-\frac{1}{\Theta_0}} 
 \exp\left(-\frac{a}{\sqrt{|N^c_{f,0} - N_f |}}\right),
\label{scaling}
\eeq  
with a scale $k_0$, the critical value $N^c_{f,0}$ of fermion flavors at
zero temperature and a (non-universal) parameter $a$. The power law 
$|N^c_{f,0} - N_f |^{-\frac{1}{\Theta_0}}$ with critical exponent $\Theta_0$ 
constitutes a universal correction to the well-known exponential Miransky 
scaling due to the running of the gauge coupling \cite{Braun:2010qs}.  
\begin{figure}[b]
\includegraphics[width=0.7\columnwidth]{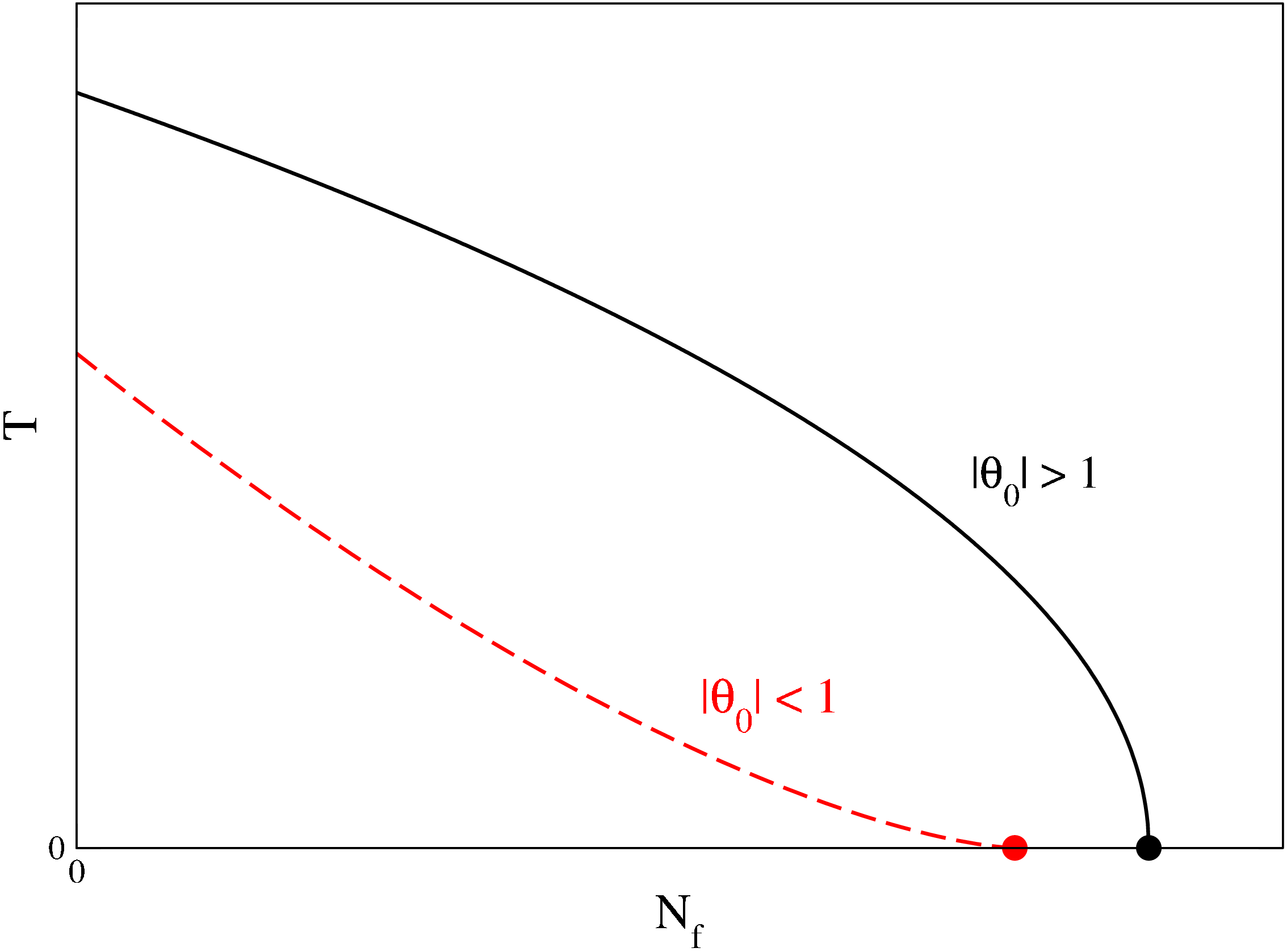}
      \caption{The generic thermal transition lines for the power law
      $|N^c_{f,0} - N_f |^{-\frac{1}{\Theta_0}}$ with cases $|\Theta_0| < 1$ and 
      $|\Theta_0 |> 1$. The quantum critical points for zero temperature are marked
      with filled circles.}
     \label{fig:generic}
\end{figure}
In principle there are two possible generic scenarios for the power law with
$|\Theta_0| > 1$ and $|\Theta_0|<1$, shown in Fig.~\ref{fig:generic}. For 
$|\Theta_0| > 1$, the phase transition line approaches the zero temperature 
axis with infinite slope, whereas for $|\Theta_0| < 1$ the slope vanishes at 
the quantum critical point. In QCD,
predictions for $\Theta_0$ from the functional renormalization group approach
\cite{arXiv:0912.4168,Braun:2010qs} are consistent with recent results
from lattice simulations \cite{arXiv:1110.3152} and result in $|\Theta_0| < 1$.

In this work we investigate the corresponding behavior of QED$_3$ under the
presence of anisotropies with $v_F = v_\Delta \neq c_s$. As a tool to determine 
the relevant critical temperatures, we use the Dyson--Schwinger equations (DSEs) 
for the (anisotropic) fermion propagator of the theory. Due to the anisotropic 
setup, the three momentum directions have to be treated separately, which prevents 
the introduction of hyperspherical coordinates in the Dyson--Schwinger equations.
It turned out that a natural framework to deal with this situation is the 
formulation of the DSEs in a finite volume, i.e. in a box with antiperiodic 
boundary conditions. This approach, originally introduced in 
Refs.~\cite{Fischer:2002eq,Fischer:2005nf}, 
has been used to study the finite volume behavior of isotropic QED$_3$ as well 
as the behavior of $N_f^c$ at zero temperature in the anisotropic setup 
\cite{Bonnet:2011hh}. The results agreed qualitatively with corresponding ones 
in other frameworks \cite{Thomas:2006bj,Concha:2009zj}, thus underlining
the feasibility of the approach. In this work we generalize the approach to
include finite temperature effects. As it turns out, we obtain values of 
$|\Theta_0|$ that depend on the size of anisotropy. 

This work is organized as follows: In the next section we discuss the structure 
of the Dyson--Schwinger equations (DSEs) of anisotropic QED$_3$ at finite temperature 
as well as the truncation scheme we are using. We also summarize some issues 
concerning the formulation of the DSEs in a box. In section \ref{results}
we present our results and discuss the scaling properties around the
quantum critical point and comment on finite volume effects. We finish with 
a short summary and conclusions.

%%%%%%%%%%%%%%%%%%%%%%%%%%%%%%%%%%%%%%%%%%%%%%%%%%%%%%%%%%%%%%%%%%%%%%%%%%%%%
\section{Technical details \label{sec:technical}}

\subsection{The Dyson--Schwinger equations in anisotropic  $\mbox{QED}_3$ 
at finite temperatures \label{subsec:anisotropicdses}}

The details of the formulation of anisotropic QED$_3$ have been discussed in
Refs.~\cite{Franz:2002qy,Lee:2002qza,Hands:2004ex} and summarized in 
Ref.~\cite{Bonnet:2011hh}. Here we only state the most essential 
details, using the notation of Ref.~\cite{Franz:2002qy}.
The fermionic four component spinors obey the Clifford algebra
$\lbrace \gamma_{\mu},\gamma_{\nu}\rbrace=2\,\delta_{\mu\nu}$.
We consider $\nf$ fermion flavors and the anisotropic fermionic velocities 
$\vf$ and $\vd$ which will be implemented via a factor $g_{i,\mu\nu}$ acting like a metric. 
The index \emph{i} = 1,2 indicates the \emph{node} in consideration. In the 
high-temperature superconducting (HTS) system, the different nodes correspond 
to different zeroes of the fermionic energy gap function placed on the Fermi 
surface of the HTS system \cite{Herbut:2002yq}.
%%%%%%%%%%%%%%%%%%%%%%%%%%%%%%%%%%%%%%%%%%%%%%%%
%%%%%%%%%%%%%%%%%%%%%%%%%%%%%%%%%%%%%%%%%%%%%%%%
\begin{figure}[t!]
		\begin{center}
			\subfigure{\label{fig:fermiondse}
                                  \includegraphics[width=0.9\columnwidth]{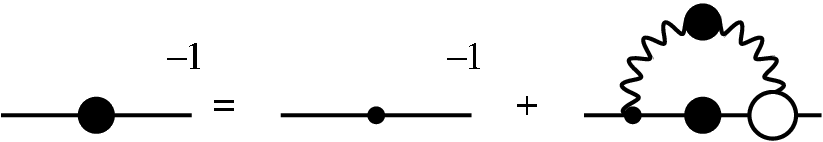}}
	                           \hspace{0.08\columnwidth}\vspace*{3mm}\\
                     
	                \subfigure{\label{fig:photondse}
                                  \hspace{-2mm}
                                  \includegraphics[width=0.9\columnwidth]{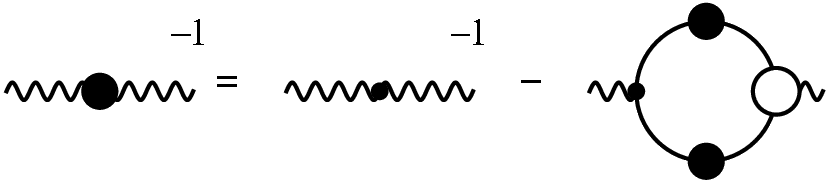}}
	                           \hspace{0.1\columnwidth}
	        \end{center}
                \caption{The diagrammatic representation of the Dyson--Schwinger equations
                         for the fermion (a) and gauge boson propagator (b).
                         Wiggly lines denote photon propagators, straight lines 
                         fermion propagators. 
                         A blob denotes a dressed propagator or vertex, whereas a 
                         dot stands for a bare fermion-photon vertex.}
\end{figure}
%%%%%%%%%%%%%%%%%%%%%%%%%%%%%%%%%%%%%%%%%%%%%%%%
%%%%%%%%%%%%%%%%%%%%%%%%%%%%%%%%%%%%%%%%%%%%%%%%

The metric for the first node is given by
\begin{equation}
\left(g_{i=1}^{\mu\nu}\right)=\left(\begin{array}{ccc}
                                        1 &    0     & 0      \\
                                        0 & \,\vf^{2}& 0      \\
                                        0 &    0     & \vd^{2}
                                  \end{array}
                            \right),
\label{eq:metric 1}
\end{equation}
whereas the metric for the second node can be obtained from interchanging the fermionic 
velocities. It enters the Lagrangian in the form of
\begin{align}
\mathcal{L}^{aniso}=\frac{N_f}{2}\sum_{j=1,2}
                         \bar{\Psi}_{j}
                         \left\{ \sum_{\mu=0}^{2}\gamma_{\nu}\sqrt{g}_{j,\nu\mu}
                                 \left(i \partial_{\mu} - a_{\mu}\right)
                         \right\}
                         \Psi_{j}.
\label{eq:lagrangian}
\end{align}
The order parameter for chiral symmetry breaking is the chiral condensate 
which can be determined via the trace of the dressed fermion propagator
\beq
S_{F,j}(\omega_p,\vec{p}) = \int^{\beta}_0 d\tau \int \frac{d^2x}{(2\pi)^2} 
e^{i\omega_p \tau- i\vec{p}\vec{x}} 
\langle \bar{\Psi}_j(\tau,\vec{x}) \Psi_j(0,0) \rangle,
\eeq
with inverse temperature $\beta=1/T$ and spacial momentum $\vec{p} = (p_1,p_2)^T$. 
The corresponding expression for the dressed photon propagator is given by
\beq
D_{\mu\nu}(\omega_p,\vec{p}) = \int_0^{\beta}d\tau\int \frac{d^2x}{(2\pi)^2} e^{i\omega_p\tau-i\vec{p}\vec{x}} 
\langle A_\mu(\tau,\vec{x}) A_\nu(0,0) \rangle,
\eeq
and the Matsubara frequencies in both equations read 
\begin{equation}
 \omega_p = 2 \pi T \left  \lbrace \begin{array}{cc}
                            n_t + \frac{1}{2} & \mbox{for  fermions}\\
                            n_t               & \mbox{for  bosons }
                         \end{array}
 \right.
\end{equation}
with integer numbers $n_t$.

Diagrammatically, the  Dyson--Schwinger equations are represented in 
Fig.~\ref{fig:fermiondse} and Fig.~\ref{fig:photondse}. 
In Euclidean space-time at  non-zero temperatures, they are explicitly  given by
\begin{widetext}
\begin{eqnarray}
S_{F,j}^{-1}(\omega_p,\vec{p})\;&=& S_{0,j}^{-1}(\,\omega_p,\vec{p}\,)
                                \;\;+\,Z_{1}\,e^{2}\,T\sum_{n_t}\int\hspace{-1mm}\frac{d^{2}q}{(2\pi)^{2}}
                                            (\sqrt{g}_{j,\mu\alpha}\gamma^{\alpha}S_{F,j}(\,\omega_q,\vec{q}\,)
                                             \sqrt{g}_{j,\nu\beta}\Gamma^{\beta}(\,\omega_q,\vec{q},\omega_p,\vec{p}\,)\,
                                              D_{\mu\nu}(\,\omega_k,\vec{k}\,)),
\label{eq:fermiondse} \\
D_{\mu\nu}^{-1}(\omega_p,\vec{p})  &=& D_{0,\mu\nu}^{-1}(\omega_p,\vec{p}\,)
                                        -\frac{Z_{1}e^{2}T N_f}{2}\hspace{-2mm}\sum_{\tiny \begin{array}{c} j=1,2;\\n_t \end{array}}\hspace{-2mm}\int\hspace{-1mm}\frac{d^{2}q}{(2\pi)^{2}}
                                         \mbox{Tr}\left[\sqrt{g}_{j,\mu\alpha}\gamma^{\alpha}S_{F,j}(\omega_q,\vec{q}\,)
                                                        \sqrt{g}_{j,\nu\beta}\Gamma^{\beta}(\omega_p,\vec{p},\omega_q,\vec{q}\,)
                                                        S_{F,j}(\omega_k\vec{k}\,)
                                                     \right]\hspace{-1mm},
\label{eq:photondse}
\end{eqnarray}
\end{widetext}
with the momentum $\,k_{\mu}\,$ defined by the difference $p_{\mu}-q_{\mu}$ and
the node index $i=1,2$. The inverse bare fermion propagator is given by
$S_{0,j}^{-1}(p) = Z_2 i \gamma_{\nu}\,\sqrt{g}_{j,\nu\mu}\, p_{\mu}$ with  
$p=(\,\omega_p,\vec{p})$ and the inverse bare photon propagator is given by
$D_{0,\mu \nu}^{-1}(p) = Z_3 p^2 P^T_{\mu\nu}(p)$ with transverse projector
$P^T$. The dressed fermion-boson vertex is denoted by 
$\Gamma^{\beta}(\,p_{\mu},q_{\mu}\;)$. The renormalization constants $Z_1$,
$Z_2$ and $Z_3$ of vertex, fermion and gauge boson respectively are each 
defined by the ratio of renormalized to unrenormalized dressing function 
of the corresponding one-particle irreducible Greens' function. Since 
QED$_3$ is ultraviolet finite, all renormalization constants can be set to 
one without loss of generality. The Ward-identity $Z_1=Z_2$ is then trivially 
satisfied.

Since we consider anisotropic space-time, we have to take care of the 
individual dressing of the Dirac components of the fermions 
at the different nodes. In order to keep track of the general structure 
of the equations we use the shorthands \cite{Bonnet:2011hh}: 
\begin{equation}
 \overline{p}_{i}^{2}:=p_{\mu}\;g_{i}^{\mu\nu}\;p_{\nu}
\label{eq:barconvention}
\end{equation}
and
\begin{align}
 \widetilde{p}_{\mu,i}:=A_{\mu,i}\left(\omega_p,\vec{p}\;\right) p_{\mu},\mbox{ (no summation convention !), }
\label{eq:tildeconvention}
\end{align}
where $A_{\mu,i}$ denotes the vectorial 
fermionic dressing function at node $i$.  

The anisotropic expressions for the dressed fermion
and photon propagators then read
\begin{eqnarray}
S_{F,i}^{-1}\left(\omega_p,\vec{p}\;\right)&=&B_{i}\left(\omega_p,\vec{p}\;\right)+\mbox{i}\;\sqrt{g_{i}}^{\mu\nu}\gamma_{\nu}\;\widetilde{p}_{\mu,i},
\label{eq:fermion}
\\
D_{\mu\nu}\left(\omega_p,\vec{p}\;\right)^{^{-1}}&=&p^{2}\left(\delta_{\mu\nu}-\frac{p_{\mu}p_{\nu}}{p^{2}}\right)
                                              +\Pi_{\mu\nu}\left(\omega_p,\vec{p}\;\right),\,\,\,\,\,
\label{eq:photon}
\end{eqnarray}
where $B_{i}$ denotes the scalar fermion dressing function at node \emph{i} and 
$\Pi_{\mu\nu}\left(\,\omega_p,\vec{p}\;\right)$ the vacuum polarization of the 
gauge boson field. 

Next we explain our approximation scheme for the DSEs of the fermion and photon
propagator. One of the most employed schemes is the $1/N_f$ expansion, which
led to some success in the early analysis of the chiral transition at large $N_f$
and zero temperature \cite{Appelquist:1988sr,Nash:1989xx}. In this approximation the vector
part of the fermion propagator is just a constant, i.e. dressing effects in the
fermion wave function are not taken into account. Later on it turned out 
that this approximation is too drastic and leads to the wrong behavior of the
photon propagator in the chirally symmetric phase. In Ref.~\cite{Fischer:2004nq} 
it was shown that a self-consistent calculation of the coupled system of fermion 
and photon-DSEs beyond the $1/N_f$ approximation leads to nontrivial power laws 
of the photon and fermion wave functions in the symmetric phase with values 
of irrational exponents depending on the details of the approximation for the 
fermion-photon vertex. A nontrivial fermion wave function beyond $1/N_f$ was
also shown to be important to correctly determine the dependencies of the
chiral transition on $N_f$ in the presence of non-zero 
anisotropies \cite{Bonnet:2011hh}. We therefore have to take into account
these effects here as well. To this end we employ the same approximation 
scheme as detailed in Ref.~\cite{Bonnet:2011hh}. We use a form for the 
fermion-photon vertex that agrees with its Ward-Takahashi identity to 
leading order in the tensor structures. This ansatz is given by
\begin{equation}
	\Gamma^\beta_i(\omega_p,\vec{p}, \omega_q,\vec{q}) = 
	\gamma^\beta \frac{A^\beta_i(\omega_p,\vec{p}) + A^\beta_i(\omega_q,\vec{q})}{2} 
	\label{vertex}
\end{equation}
without using summation convention and the fermionic momenta $p,q$ 
at the vertex. In the gauge sector we are currently not able to solve 
the photon DSE self-consistently for general anisotropies. We therefore
approximate the photon self-energy by a model which is closely built along
the self-consistent result of the isotropic case. In particular it takes 
into account the irrational power laws discovered in Ref.~\cite{Fischer:2004nq}. 
The gauge boson vacuum polarization then reads    
\begin{eqnarray}
\Pi^{\mu\nu}\left(p\;\right) & =&  \sum_{i}\sqrt{\overline{p}_i^2}
\left(g_{i}^{\mu\nu}-\frac{g_{i}^{\mu\alpha} p_{\alpha}\;g_{i}^{\nu\delta}p_{\delta}}
{\overline{p}_i^2}\right)\,\,
\Pi_i\left(p\right)\,,\nonumber\\
	\Pi_i\left(p\right) &=& \frac{e^2 N_f}{16\,v_{F}v_{\Delta}}  
	\left(\frac{1}{\sqrt{\overline{p}_i^2}}\frac{\overline{p}_i^2}{\overline{p}_i^2+e^2} 
	+ \frac{1}{\overline{p}_i^{1+2\kappa}}\frac{e^2}{{\overline{p}_i}^2+e^2}\right)\,,\nonumber\\
	&& \label{model}
\end{eqnarray}
with  $\overline{p}_{i}^{2}$ defined in Eq.~(\ref{eq:barconvention}) and 
$p=(\,\omega_p,\vec{p})$. In Ref.~\cite{Fischer:2004nq} the value of
$\kappa$ for the vertex truncation Eq.~(\ref{model}) at the critical number of
fermion flavors $N_f^c$ has been determined as $\kappa=0.0358$. In principle, 
this number may depend on the fermion velocities in the anisotropic 
case. Nevertheless, due to our current inability to calculate this dependence
we will keep $\kappa$ constant in all calculations. We also fix the gauge coupling
$e^2=1$ independent of the value of $N_f$ and temperature $T$. This choice is
one of many possible ones that serve to compare theories with different numbers
of fermion flavors $N_f$. 

Having specified our approximations (\ref{vertex}),(\ref{model}), we now proceed 
with the fermion-DSE.
In order to extract the scalar and vector dressing function of the fermion
we take appropriate traces of the fermion-DSE and rearrange the equations 
to end up with

\begin{widetext}
\begin{eqnarray}
%
%%Fermion scalar dressing
B_{i}\left(\omega_p,\vec{p}\;\right)\hspace{-1mm}&   = &\hspace{-1mm} \; Te^{2}\sum_{n_t}\int\hspace{-1mm}\frac{d^{2}q}{\left(2\pi\right)^{2}}
                                        \frac{B_{i}\left(\omega_q,\vec{q}\;\right)g_{i}^{\mu\nu}D_{\mu\nu}(\omega_k,\vec{k}\;)}
                                             {B_{i}\left(\omega_q,\vec{q}\;\right)^{2}+(\overline{\widetilde{q}}_{i,\mu}\;)^{2}},
\label{eq:bdse}
                                 \\
%
%%Fermion vector dressing
A_{\mu,i}\left(\omega_p,\vec{p}\;\right)\hspace{-1mm} &  = &\hspace{-1mm} 1 -\frac{Te^{2}}{p_{\mu}}\hspace{-1mm}\sum_{n_t}\hspace{-1mm}
                                        \int\hspace{-2mm}\frac{d^{2}q}{\left(2\pi\right)^{2}}\hspace{-1mm}
                                        \left\{ \hspace{-1mm}
                                            \frac{2\left(\widetilde{q}_{\lambda,i}\;g_{i}^{\lambda\nu}D_{\mu\nu}  (\omega_k,\vec{k}\;)\right)
                                                        -\widetilde{q}_{\mu,i}\;      g_{i}^{\lambda\nu}    D_{\lambda\nu}(\omega_k,\vec{k}\;)}
                                                 {B_{i}\left(\omega_q,\vec{q}\;\right)^{2}+(\;\overline{\widetilde{q}}_{i,\mu}\;)^{2}}\hspace{-1mm}
                                        \right\} \hspace{-1mm}
\frac{A_{\mu,i}(\omega_p,\vec{p}) + A_{\mu,i}(\omega_q,\vec{q})}{2},               
\label{eq:adse}                  
\end{eqnarray}
\end{widetext}
where again we have no summation for the external index $\mu$.
In the following, we will investigate the finite temperature equations in case 
of equal fermionic velocities $\vf = \vd$, reducing the fermionic vector dressing function 
to two instead of three different components and leaving us with only one sort of nodes. 
We therefore drop the nodal index $i$. 

\subsection{The DSEs on a torus \label{subsec:dsetorus}}

Considering anisotropy and additional finite temperatures, the computational 
demands increase significantly in comparison to the isotropic vanishing temperature 
investigations, even despite the applied approximations for the gauge boson.  
We therefore continue our investigations by changing the basic manifold from 
three-dimensional Euclidean spacetime to a three-torus, which is well suited for the 
evaluation of  the Cartesian sums emerging from the boundary conditions the 
fermion and boson fields have to obey.  
To be precise, only the time direction for the boson has to be periodic and for 
the fermion field antiperiodic. The spacelike components 
are chosen to show the same (anti-)periodicity as the time component of the according 
field for convenience. 
In order to guarantee a valid interpretation of the timelike direction as a temperature, 
we have to make sure that the aspect ratio of time- and spacelike 'boxlength' is appropriate.   
We therefore distinguish the boxlength in coordinate space as $L_1 \sim 1/T$ in time direction and 
equal lengths $L_2 =L_3 = L_X$ in spacelike directions. 
The momentum integrals are then also replaced by Matsubara sums, 
\begin{equation}
\int \frac{d^2q\,}{(2 \pi)^2} \:(\cdots) \:\:  \longrightarrow \:\:\frac{1}{L_X^2}
\sum_{n_2,n_3} \:(\cdots) \,. 
\end{equation}
The discrete fermionic momenta are then counted by
${\bf q}_{\bf n} = \sum_{i=2,3} (2\pi/L_X)(n_i+1/2) \hat{e}_i$,
where $\hat{e}_i$ represents a Cartesian unit vector in Euclidean 
momentum space. For the
photon with periodic boundary conditions the momentum counting goes like 
${\bf q}_{\bf n} = \sum_{i=2,3} (2\pi/L_X)(n_i) \hat{e}_i$.
The resulting system of equations can be solved numerically along the lines
described in Ref.~\cite{Goecke:2008zh}. 

As already discussed in the course of previous investigations 
\cite{Bonnet:2011hh,Goecke:2008zh}, evaluations on a finite volume 
have a considerable impact on the quantitative results for the 
critical number of fermion flavors, but hardly change the qualitative 
behavior, when considering anisotropies. Since the introduction of finite 
temperature is conceptually very similar to the previously studied 
anisotropies, we expect similar findings for volume extrapolations and 
postpone detailed studies on this subject. Instead, we concentrate again 
on the variations of $N_f^c $, keeping in mind that its absolute number 
will change in infinite volume calculations.

%%%%%%%%%%%%%%%%%%%%%%%%%%%%%%%%%%%%%%%%%%%%%%%%%%%%%%%%%%%%%%%%%%%%%%%%%%%%%
\section{Numerical results \label{results}}

\subsection{The critical $\mbox{N}_{f}^{c}$ in the anisotropic case $\vf = \vd$ \label{subseq:phasediagrams}}

\begin{figure}[t]
\begin{center}
\includegraphics[width=0.9\columnwidth]{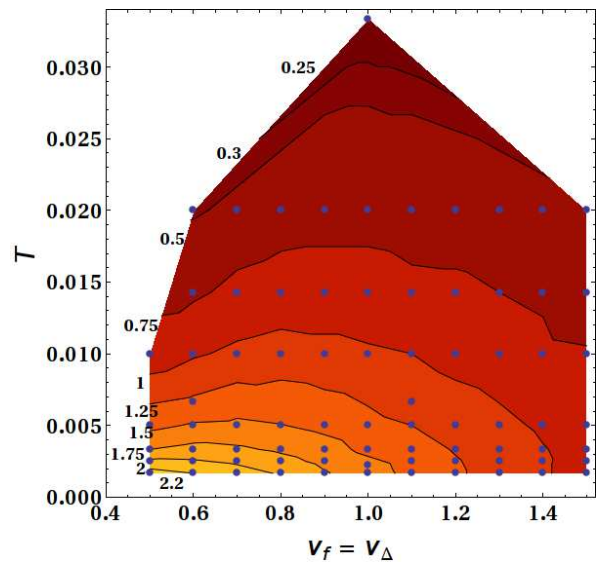}
\end{center}
      \caption{The critical number of fermion flavors, $N_{f}^c$ plotted as a 
      function of the fermionic velocities $\vf = \vd$, and the temperature $T$.
      \label{fig:phasediagram}}
\end{figure}

In this section we present our results on the finite temperature phase diagram for 
the case of equal anisotropic velocities $\vf = \vd $ with $c_s=1$. 
We obtain the phase diagram for the critical number of fermion flavors, $N_f^c$, 
from solving the set of Dyson--Schwinger equations (\ref{eq:bdse}) and (\ref{eq:adse})
on a torus of $(N_X^2 = 39^2) \times (N_T=23)$ points and a box length of 
$e^2 L_X  = 600$. In time direction, the sum of $N_T= 23 $ Matsubara frequencies  
guarantee a reasonable aspect ratio between the spacial and temporal number of 
lattice points, necessary for finite temperature calculations. The order parameter
for chiral symmetry breaking is the chiral condensate ($tr_D$ denotes the trace in
Dirac-space),
\beq
\langle \bar{\Psi} \Psi \rangle = T \sum_{n_t} \int \frac{d^2q}{(2\pi)^2}
tr_D S_F(\omega_q,\vec{q})
\eeq
or, equivalently, the value of the scalar fermion dressing function 
$B(\omega_q,\vec{q})$ at the lowest Matsubara frequency and two-momentum. 
For each given temperature and anisotropy we determined the critical number of
fermion flavors $N_f^c$.

The results of our calculations are shown in Fig.~\ref{fig:phasediagram}. 
The critical number of fermion flavors has been determined numerically at each dot 
in the diagram. Subsequently a Mathematica interpolation routine converted these 
values into areas of equal $N_f^c$ in different colors separated by contour lines.
The contour lines therefore have to be taken as approximate results from the 
interpolation procedure and only serve to guide the eye. In agreement with our 
previous investigations \cite{Bonnet:2011hh} we find a decreasing critical 
number of fermion flavors for increasing anisotropy at effective zero temperature. 
When we increase the temperature at fixed anisotropy we again find a decreasing 
$N_f^c$ for all anisotropic velocities taken into account.  
It is noteworthy that for anisotropic velocities larger than the isotropic value, 
$\vf = \vd \geq 1$, this decrease happens at a comparably slow rate, while for 
$\vf = \vd \leq 1 $ the critical number of fermion flavors rapidly falls off 
with increasing temperatures. 

\begin{figure}[t]
\begin{center}
\includegraphics[width=0.9\columnwidth]{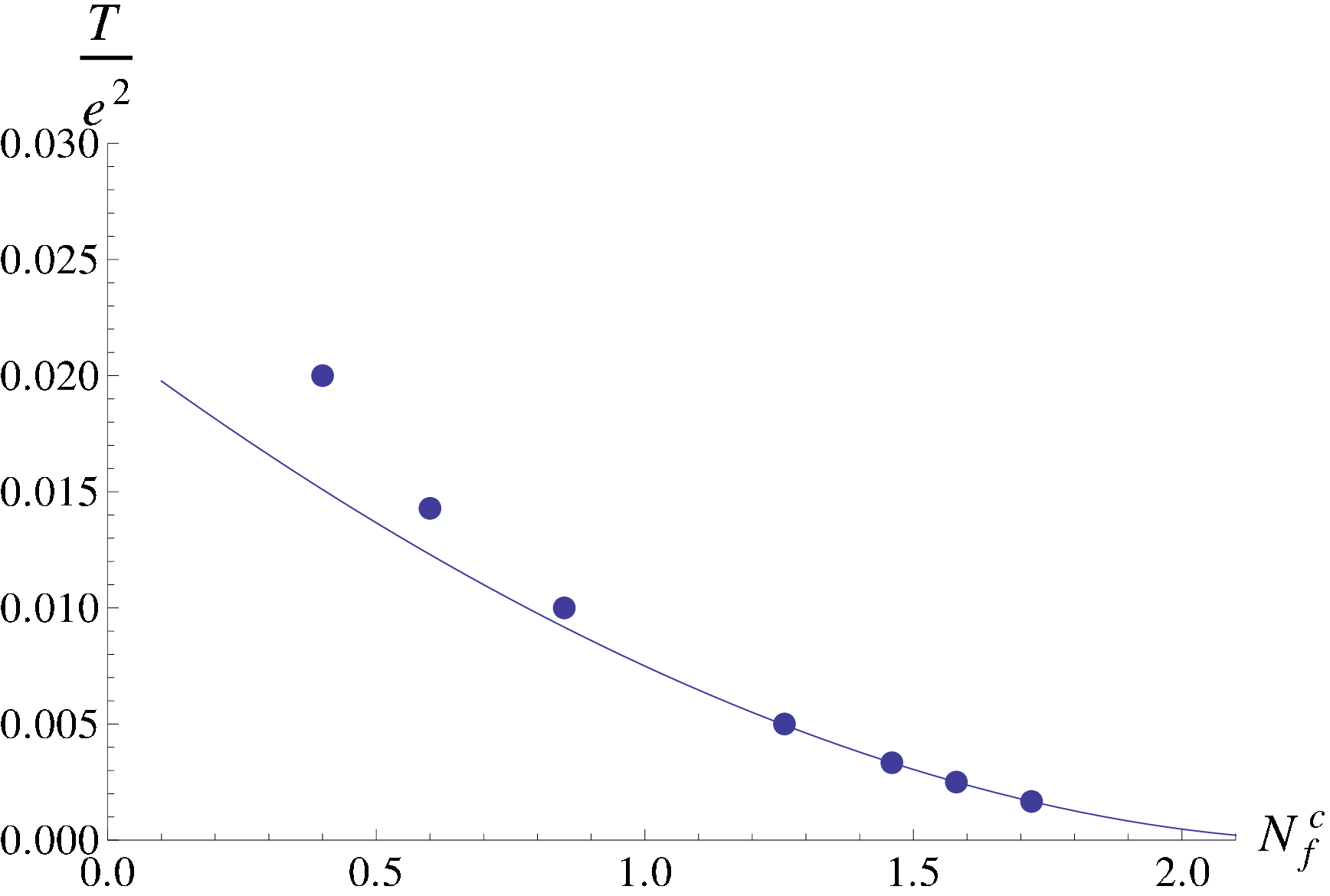}
\end{center}
      \caption{The critical temperature for the chiral transition as a function of
      the number of fermion flavors for the case $\vf = \vd = 0.8$. Shown are our
      numerical results together with a fit of the power law (\ref{eq:tcr}) in the
      region $N_f > 1.25$, i.e. in the vicinity of the quantum transition.}
    \label{fig:v08}
\end{figure}

\subsection{ Scaling in the pseudo-conformal phase transition region
\label{subsec:infinite volume limit}}

There is a considerable number of works identifying the 
chiral phase transition in QED$_3$ as of pseudo-conformal nature 
due to the dimensionful coupling constant $\alpha$, see e.g. 
\cite{Miransky:1988gk,Gusynin:1998kz,Fischer:2004nq} and references therein.
The scaling behavior close to the phase transition relates QED$_3$ with other 
strongly coupled theories such as many-flavor QCD that has been subject to 
a number of recent studies in the framework of renormalization group equations 
\cite{Gies:2005as,Braun:2005uj,Braun:2006jd,arXiv:0912.4168,Kaplan:2009kr,Braun:2010qs,Braun:2011pp}.
These investigations have shown that the explicit form of the 
symmetry-breaking scale, $k_{SB}$, is to zeroth order given by the expression 
\begin{eqnarray}
 k_{SB} &\sim& k_{0} \theta(N^c_{f,0}-N_f) 
 |N^c_{f,0} - N_{f}|^{-\frac{1}{\Theta_0}} \nonumber\\
            &       &\times\mbox{exp}\left(-\frac{a}{\sqrt{|N^c_{f,0}- N_f |} } \right),
\label{eq:ksb}
\end{eqnarray}
where $k_0$ denotes a scale-fixing constant and $N^c_{f,0}$ is the critical fermion 
number for chiral symmetry breaking at zero temperature. Furthermore, 
$\Theta_0$ represents the leading order term in 
the Taylor expansion of the critical exponent $\Theta$ in $|N_{f,0}^c-N_f|$ and 
$a$ a further parameter. Whereas the form of the scaling relation (\ref{eq:ksb}) 
has been derived from very general considerations and is therefore universal for 
strongly flavored asymptotically free gauge theories, the parameter $a$ depends 
on the theory in question, see Ref.~\cite{Braun:2010qs} for details. A similar 
type of scaling law also applies to observables such as the critical temperature. 

In a formulation of the theory on a finite volume it is extremely difficult to 
assess the complete scaling law (\ref{eq:ksb}). Exponential Miransky scaling 
dominates (\ref{eq:ksb}) only very close to the critical fermion 
number and therefore cannot be resolved within the temperature range available in 
such a setup. This is also the case for our formulation of the DSEs on a 
torus\footnote{The reason is a matter of scales. Close to $N_{f,0}^c$ there is
a large separation of the dynamical scale, i.e. the generated fermion 
mass, and the intrinsic scale given by the dimensonful coupling $\alpha$ of 
QED$_3$. This separation of scales necessitates extremely large volumes which 
requires a continuum formulation of the DSEs \cite{Goecke:2008zh}. 
Lattice calculations also have this problem.}. 
In the following we therefore strive to identify the power law part of the scaling 
relation, which is expected to dominate further away from $N_{f,0}^c$ until finally
a value of $|N_{f,0}^c-N_f|$ is reached where scaling ceases altogether. In the
following we seek to locate the region where the power law scaling  
\begin{equation}
 T_{cr} \sim k_0 |N^c_{f,0} - N_f |^{-\frac{1}{\Theta_0}}
\label{eq:tcr}
\end{equation}
is applicable\footnote{Strictly speaking, the relation (\ref{eq:tcr}) is only an 
upper bound for the chiral phase transition temperature since it is insensitive 
to local ordering phenomena due to Goldstone modes in the deep infrared 
\cite{Braun:2010qs,Braun:2009si}, which may yield corrections to (\ref{eq:tcr}). 
However, since in our formulation at finite volume and small lattice sizes such 
corrections cannot be resolved we stick to (\ref{eq:tcr}) as a good approximation.}.
We come back to this point below.

As an example consider the case of $\vf=\vd=0.8$ shown in Fig.~\ref{fig:v08}. The
power law (\ref{eq:tcr}) nicely fits the calculated points in the region 
$1.25 < N_f < 1.75$. For $N_f < 0.9$ significant deviations of the calculated 
critical temperatures from the scaling law occur indicating the finite extent
of the scaling region. For $N_f > 1.75$, somewhere close to $N^c_{f,0}$ we 
expect the exponential Miransky scaling finally to dominate over (\ref{eq:tcr}).
This region can only be investigated in the (even more demanding) continuum 
formulation of the DSEs, which we will address in future work. 

\begin{figure}[t]
 \begin{center}
\includegraphics[width=0.9\columnwidth]{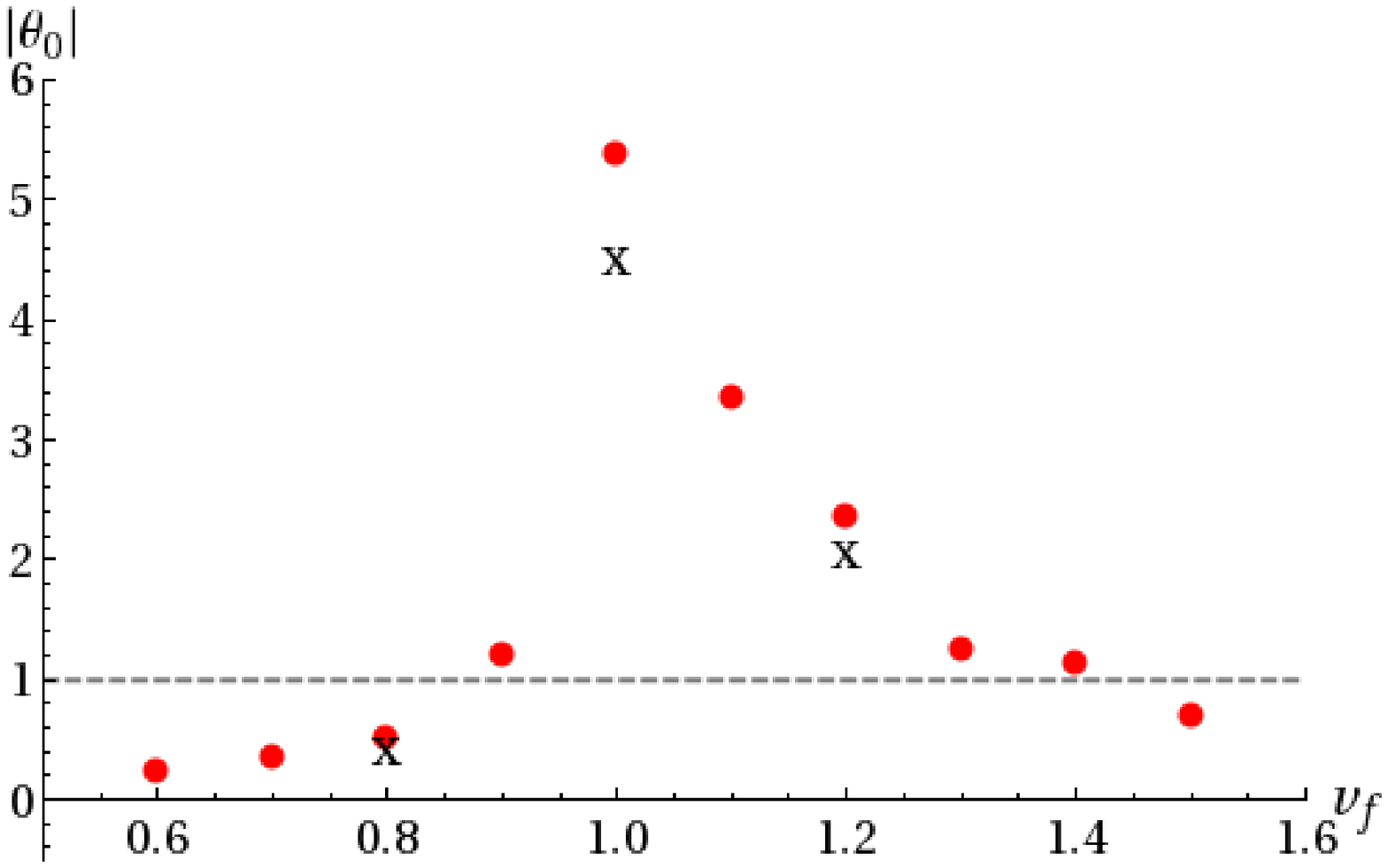}
\end{center}
      \caption{The critical exponent $|\Theta_0|$ plotted as a 
      function of the fermionic velocities $\vf = \vd$. Red dots 
      indicate results obtained with a box length of $L_X=600/e^2$, 
      whereas the crosses indicate results on the larger torus
      with $L_X=900/e^2$}
     \label{fig:theta}
\begin{center}
\includegraphics[width=0.9\columnwidth]{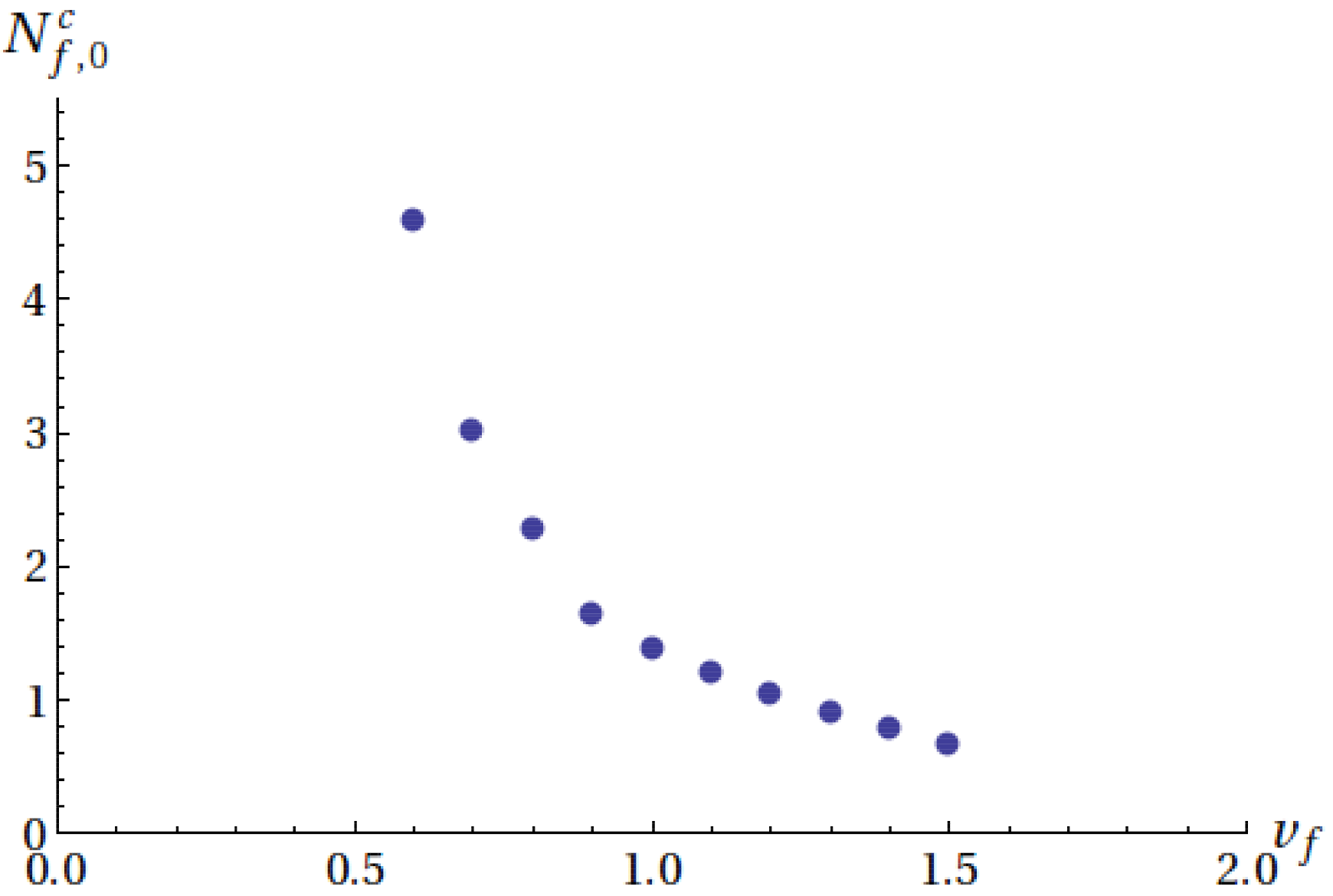}
\end{center}
      \caption{The critical number of fermion flavors for zero 
      temperature, $N^c_{f,0}$ plotted as a 
      function of the fermionic velocities $\vf = \vd$.}
    \label{fig: nf0c}
\end{figure}

The values for $|\Theta_0|$ and the corresponding critical fermion 
numbers $N_{f,0}^c$ at zero temperature are given in Figs.~\ref{fig:theta} 
and \ref{fig: nf0c}. In general we observed that the size of the scaling 
region where the power law (\ref{eq:tcr}) can be found, depends on the value 
of $|\Theta_0|$. It is quite large for $|\Theta_0| <1$, as can be seen in 
Fig.~\ref{fig:v08}, but turned out to be much smaller for $|\Theta_0| >1$ 
and consequently the extracted values for $\Theta_0$ in this case have to 
be treated with some caution. 

Fig.~\ref{fig:theta} shows that the anisotropic velocities provide a parameter 
that qualitatively changes the scaling behavior close to the quantum critical 
point $N_{f,0}^c$. Shown are results for our standard torus of 
$(N_X^2 = 39^2) \times (N_T=23)$ points and a spatial box length of 
$e^2 L_X = 600$, denoted by red dots. In order to assess the effects of finite 
volume we also determined three selected points (the crosses) on an even 
larger torus with $(N_X^2 = 59^2) \times (N_T=23)$ and $e^2 L_X  = 900$, 
requiring a considerable increase in invested computational resources.
Finite volume effects are clearly present and tend to amplify the results for 
the critical exponent. However, this quantitative impact is rather uniform with 
respect to changes in the anisotropy, a finding in agreement with similar estimates 
in the zero temperature case\cite{Bonnet:2011hh}\footnote{In principle, also finite 
size effects are possible in a torus calculation. In QED$_3$, however, the 
dimensionful coupling serves as a natural cutoff for large momenta and therefore 
finite size effects are small once the lattice spacing is smaller than this scale. 
In our calculation this is the case.}. 
We therefore conclude that our results  serve very well to identify at least the 
qualitative behavior of $|\Theta_0|(v_f)$. We find a maximum with a critical 
exponent much larger than one in the isotropic case $\vf = \vd = 1 $. For smaller 
anisotropies, $0.5 <\vf = \vd < 0.9$, the critical exponent decreases rapidly to 
values smaller than one, $|\Theta_0| < 1$, while for larger anisotropies 
$1.1 < \vf = \vd < 1.5 $ the decrease is slower. 

The size of the exponent $\Theta_0$ is a measure for the size of the power law 
corrections to exponential Miransky scaling close to $N_{f,0}^c$, as discussed 
in Ref.~\cite{Braun:2010qs}. These corrections are induced by the running 
of the gauge coupling determined by the photon polarization Eq.~(\ref{model})
and found to be small for $|\Theta_0| \gg 1$ and large for $|\Theta_0| \ll 1$.
Thus the large value of $|\Theta_0| \approx 5$ for the isotropic case
indicates the presence of dominant exponential Miransky-scaling with only
very small power law corrections. On the other hand, for anisotropies 
$\vf = \vd < 0.9$ the power law corrections dominate. Unfortunately at present 
we cannot further quantify this issue since the exponential Miransky scaling
region very close to $N_{f,0}^c$ is typically small (see Table 2 in 
Ref.~\cite{Braun:2011pp} for an estimate)
and therefore not accessible in our formulation on a 
torus, as discussed above. Nevertheless we may be able to
qualitatively infer the size of the scaling region by going further away
from $N_{f,0}^c$. Indeed, as discussed above, we find a larger
scaling region there for $|\Theta_0| < 1$, whereas for $|\Theta_0| > 1$ the 
scaling region is much smaller in agreement with the considerations above
and in Ref.~\cite{Braun:2010qs}. 

The zero-temperature critical number of fermion flavors (Fig.~\ref{fig: nf0c}) 
decreases for increasing anisotropies thus confirming the behavior already 
found in Ref.~\cite{Bonnet:2011hh}. Note that the zero-temperature fit 
of the critical number of fermion flavors is not identical to our previous 
results. The (small) deviations arise from the fact, that both calculations 
were performed on similar, but not equally sized tori.

\section{Summary and conclusions} \label{summary}

In this work we have determined the critical temperature for the chiral phase 
transition of anisotropic QED$_3$ as a function of the number of fermion flavors 
$N_f$ for different anisotropies. To this end we solved a coupled set of 
truncated DSEs for the anisotropic fermion propagator together with model 
input for the photon self energy determined from the isotropic case at $T=0$. 
We found a generic phase diagram with decreasing critical temperature with 
increasing $N_f$. The transition line hits the $N_f$-axis at the quantum 
critical point $N_{f,0}^c$. We investigated the critical scaling with $N_f$ 
around the critical point and verified a universal scaling law determined on
general grounds for strongly coupled asymptotically free gauge theories in 
Refs.~\cite{arXiv:0912.4168,Braun:2010qs}. A characteristic 
quantity that characterizes this scaling region is the critical exponent 
$|\Theta_0|$. We found that the anisotropic velocities provide a parameter 
that drives QED$_3$ from a setting with $|\Theta_0|$ smaller than one to 
a region with large values of the critical exponent $|\Theta_0| \gg 1$. 
This result is interesting but preliminary, since the gauge sector of the
theory was not taken into account dynamically in our framework. Thus the 
full consequences of our findings can only be addressed when fully 
self-consistent solutions of the fermion- {\it and} photon-DSEs are 
available. Nevertheless, our present framework serves as a qualitative 
guide and constitutes a proof of principle for the feasibility of dealing with 
anisotropic systems in strongly interacting fermionic theories.

%%%%%%%%%%%%%%%%%%%%%%%%%%%%%%%%%%%%%%%%%%%%%%%%%%%%%%%%%%%%%%%%%%%%%%%%%%%%%
\section*{Acknowledgement}

We thank Jens Braun, Holger Gies and Richard Williams for many fruitful discussions.
This work was supported by the Helmholtz-University Young Investigator Grant 
No.~VH-NG-332,  by the Deutsche Forschungsgemeinschaft through SFB 634  and the Helmholtz
International Center for FAIR within the LOEWE program of the State of Hesse.

%%%%%%%%%%%%%%%%%%%%%%%%%%%%%%%%%%%%%%%%%%%%%%%%%%%%%%%%%%%%%%%%%%%%%%%%%%%%%

\end{document}